\documentclass[a4paper,11pt]{article}
\pdfoutput=1 

\usepackage{jheppub} 

\usepackage{amsmath, amsthm, amssymb}
\usepackage{dsfont}
\usepackage{wasysym, verbatim} 
\usepackage{color}
\usepackage{tensor}
\usepackage{mathtools}
\usepackage{enumitem}
\usepackage{physics}
\usepackage{subcaption}
\usepackage{graphicx}
\usepackage[utf8]{inputenc}
\usepackage{booktabs,multirow, quotes,tablefootnote}
\usepackage{nicematrix}
\usepackage{tikz}

\usepackage{cancel}
\usepackage[normalem]{ulem}

\usepackage{braket}
\usepackage{tcolorbox}
\usepackage{float}

\usepackage{notoccite} 

\usepackage{cleveref}
\creflabelformat{equation}{(#2\textup{#1}#3)}
\crefname{equation}{Eq.}{Eqns.}
\crefname{figure}{Fig.}{Figs.}
\crefname{section}{Section}{Sections}
\crefname{table}{Table}{Tables}
\crefname{chapter}{Chapter}{Chapters}
\crefname{appendix}{Appendix}{Appendices}
\crefname{subsection}{Section}{Sections}
\crefname{remark}{Remark}{Remarks}
\crefname{footnote}{footnote}{footnotes}

\usepackage{xcolor}
\usepackage{tcolorbox}
\newtcolorbox{mybox}{colback=gray!30,
	boxrule=0pt,arc=0pt,boxsep=2pt,left=2pt,right=2pt,leftrule=0pt}

\usepackage{hyperref}

\numberwithin{equation}{section}

\theoremstyle{theorem}
\newtheorem*{conjecture*}{Conjecture}

\numberwithin{theorem}{section}
\numberwithin{definition}{section}
\numberwithin{lemma}{section}
\theoremstyle{remark}
\newtheorem{remark}{Remark}\numberwithin{remark}{section}

\def\eps{\epsilon}
\newcommand{\beq}{\begin{equation}} 
\newcommand{\eeq}{\end{equation}}

\def\ge{\geqslant}
\def\le{\leqslant}

\def\leq{\leqslant}
\def\<{\langle}
\def\>{\rangle}

\newcommand{\myinclude}[2][]{\raisebox{0.6ex}{\raisebox{-0.5\height}{\includegraphics[#1]{#2}}}}

\usepackage{enumitem}

\renewcommand{\ge}{\geqslant}
\renewcommand{\le}{\leqslant}

\makeatletter
\def\@fpheader{\ }
\makeatother

\title{Conformal bootstrap: from Polyakov to our times}

\author{Slava Rychkov}

\affiliation{Institut des Hautes \'Etudes Scientifiques, 91440 Bures-sur-Yvette, France}

\emailAdd{slava@ihes.fr}

\abstract{We trace the history of conformal bootstrap from its early days to our times - a great example of unity of physics. We start by describing little-known details about the origins of conformal field theory in the study of strong interactions and critical phenomena in the 1960s and 1970s. We describe similarities and differences between approaches and results of the main groups in Moscow, Rome, and Sofia. Then come the breakthroughs in the 1980s and the 1990s, in particular 2D CFT and holography. Finally, we describe the genesis of the numerical conformal bootstrap, from the conformal technicolor bounds in the 2000s, to the determination of the 3D Ising critical exponents in the 2010s. We conclude with some outstanding challenges. We stress that conformal invariance is a symmetry of nature.
 }

\begin{document}

\maketitle

\flushbottom

\newpage

\section{Introduction}

After almost 20 years of conformal bootstrapping, I accumulated some interesting historical facts about our field, which I would like to share. This especially concerns the earlier history of conformal field theory in the 1960s and 1970s (Sections \ref{sec:1970}-\ref{sec:mack-and-the-sofia-group}). In the 2000s and 2010s I had the fortune to participate in a couple of memorable collaborative efforts in the numerical conformal bootstrap---the story which I will also describe (Sections \ref{sec:numCB}-\ref{sec:2011-2014---attack-on-the-3d-ising-model}). To keep from being purely historical, I included Section \ref{sec:2025---selected-challenges-for-the-conformal-bootstrap} about open problems and future directions. In the conclusions, I emphasize that conformal invariance should be rightly considered a symmetry of nature, although more experimental tests are always welcome.

\section{1970 - Migdal and Polyakov meet Kastrup in Kyiv}
\label{sec:1970}
In 1970, the XV International Conference on High Energy Physics (ICHEP) was held in Kyiv, then the capital of the Soviet Ukraine. This was the time of the iron curtain. Most Soviet scientists were not allowed by the Soviet authorities to travel abroad, and visits by the Western scientists into the USSR were rare and difficult to arrange.\footnote{For a glimpse, see an interview with Elias M. Stein, my Princeton mathematics advisor, about his 1976 visit to the USSR to meet with Gelfand \cite{Stein-Gelfand}.} So the Kyiv conference was one of the rare occasions where Western and Soviet physicists could meet. Among the attendees there was Hans Kastrup, a physicist from West Germany. Kastrup (born 1934) was among the first physicists who got interested in the conformal symmetry after World War II. In 1962, he defended PhD thesis on conformal symmetry in particle physics. Among other things, he introduced the term “special conformal transformation” \cite{Kastrup:1962zza}. Later on, he advised the PhD thesis of Gerhard Mack (1940-2023, PhD 1967). 

In Kyiv, Kastrup met two young Soviet physicists Sasha Migdal (born 1945) and Sasha Polyakov (born 1945).\footnote{Migdal and Polyakov's names are not in the list of conference participants \cite{Shelest:1972zia}, but their questions after other people's talks were recorded, see below.} Kastrup recalls that their discussions then continued in Moscow \cite[Ref.~215]{Kastrup:2008jn}. This encounter turned out to be fateful for the history of conformal field theory.

At the time, one idea in particle physics was that hadronic interactions may be asymptotically scale invariant at high energies. This was first proposed in Mack's thesis \cite{Mack:1968zz}. This idea motivated Ken Wilson's operator product expansion \cite{Wilson:1969zs}, and it was discussed at the Kyiv conference in the rapporteur talk "Chiral algebra" by Bruno Zumino \cite[p.~496]{Shelest:1972zia}. 

Migdal and Polyakov were interested not just in particle physics but also in critical phenomena.\footnote{Like Ken Wilson in the West, who did not attend the Kyiv conference.} 
Following in the footsteps of Patashinskii and Pokrovskii \cite{Patashinskii1964} and of Vaks and Larkin \cite{Vaks1965}, by 1970 Migdal and Polyakov completed a series of works on scale invariance and the emergence of non-canonical scaling dimensions in critical phenomena \cite{Polyakov1968,Polyakov1969,Migdal1968,Migdal1970}. Migdal also collaborated with Gribov on anomalous scaling in Regge theory \cite{Gribov:1968fg}.\footnote{English translations and Russian originals of all Sov.~Phys.~JETP articles can be freely downloaded at \url{http://jetp.ras.ru}.}

The Kyiv conference proceedings \cite{Shelest:1972zia} recorded questions asked after each rapporteur talk, along with the answers by the speaker and the audience. Surprisingly, when Migdal and Polyakov brought up the relevance of scale invariance for critical phenomena after Zumino's talk, this provoked hostile reaction of C.N.~Yang and T.T.~Wu:

\begin{mybox}\small
{\it Polyakov:}
I should like to point out that things similar to the Wilson expansion, scale invariance, and anomalous dimensions in the field theory were investigated several years ago in connection with critical phenomena problems. These investigations have shown that logarithmic terms in the perturbation theory, which violate scale invariance, finally sum up to give power functions of distances and hence anomalous dimensions of fields. $\langle\ldots\rangle$

{\it Yang:}
In statistical mechanics in recent years there have been many discussions of scale invariance.
These discussions were very stimulating. But I disagree with a previous comment by Dr.~Polya­kov. I do not believe that there are either mathematical reasons or physical insight that would conclusively lead to scale invariance.

{\it T. T. Wu:}
I would like to go slightly further than Professor Yang. Not only are the so-called scaling laws in statistical mechanics not well-established, there are now experimental evidences and theoretical arguments against them. The situation was clearly presented almost a year ago by Barry McCoy in Physical Review Letters.
\end{mybox}

As this exchange shows, the status of scale invariance, let alone conformal invariance, was far from clear to the community in 1970, including some of its most illustrious members. 

\section{1970 - Polyakov argues that critical phenomena are conformally invariant}
\label{sec:Polyakov1970}
On October 26, 1970, Polyakov submitted to JETP Letters a short paper "Conformal symmetry of critical fluctuations" \cite{Polyakov:1970xd}. In this paper he did three things:
\begin{enumerate}
\item[1)]
derived the expressions for 3- and 4-point functions from conformal invariance;
\item[2)]
argued that correlation functions at the critical point should be conformally invariant;
\item [3)] 
checked the predictions against the 2D Ising model.
\end{enumerate}
The results of point 1) are classic. I am pasting here the equations taken from Polyakov's paper, in self-explanatory notation: 
\beq
\myinclude{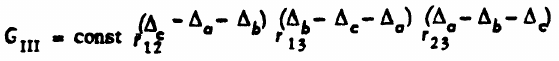}\label{eq:3pt}
\eeq
\beq
\myinclude{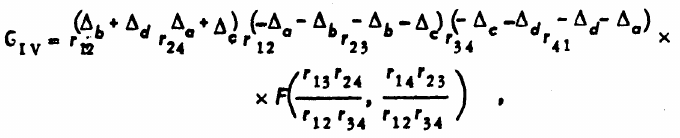}
\eeq

Points 2) and 3) are not as well known so I'd like to mention what he did. The check in point 3) was against the correlator $\langle \sigma \sigma \epsilon\rangle$ in the critical 2D Ising, extracted from the results of Kadanoff \cite{KadanoffPhysRev1969}. That was a solid check.\footnote{Ref.~\cite{KadanoffPhysRev1969} computed the critical $2n$-point functions of $\sigma$ fields on a line, and then used OPE to extract $n$-point functions of $\epsilon$ fields on a line. This and \cite{KadanoffPRL1969} are some of the earliest applications of OPE. Correlator $\langle \sigma \sigma \epsilon\rangle$ for three points on a line is not given by Kadanoff, but can be easily extracted using the OPE.}

Polyakov's argument in point 2)~may not appear convincing to a modern reader, but it's interesting from the historical perspective and to understand \cref{sec:old-bootstrap} below.
It was based on partially resummed perturbation theory, in terms of exact propagators and exact vertex functions. This was one approach to critical phenomena under development in the Soviet Union, within which some toy-model calculations could be done, using various approximations \cite{Patashinskii1964,Polyakov1968,Gribov:1968fg,Migdal1968}. One gets a diagram technique, sometimes called "skeleton expansion", with a smaller number of graphs than the full set of Feynman diagrams. Polyakov observed that conformal invariance assumption is self-consistent for the skeleton expansion: if one uses conformal $2$-point functions as propagators and (amputated) conformal $n$-point functions as vertices, then all skeleton graphs respect conformal invariance.

In the acknowledgments of \cite{Polyakov:1970xd}, Polyakov thanked "H. Kastrup (West Germany) for explaining the mathematics of the conformal group."

\begin{remark}
	Simultaneously\footnote{Schreier's paper was submitted to a journal a couple of months before but appeared in print a couple of months after Polyakov's work.} with Polyakov, the problem of conformal three-point functions was considered by Ethan J. Schreier \cite{Schreier:1971um}, PhD student of Kenneth Johnson at MIT.  Schreier's paper focused on vector and axial currents in a parity-invariant 4D CFT. As such it's the first work which imposed constraints of conformal invariance on correlators of operators with spin. Schreier's work is motivated by the desire to understand the axial anomaly. There is no connection to critical phenomena. Currents are assumed conserved at non-coincident points and have canonical dimension.\footnote{Some information about subsequent Schreier's career in experimental astrophysics can be found in \url{https://spacenews.com/ethan-j-schreier-to-become-president-of-associated-universities-inc/}}
	\end{remark}
	
\section{In search of a dynamical principle - ``old bootstrap''}
\label{sec:old-bootstrap}

Polyakov's 1970 paper \cite{Polyakov:1970xd} was a crucial step towards establishing conformal symmetry in the theory of critical phenomena. But how does it help to compute the critical exponents? The first idea was to used the skeleton expansion, as it was self-consistent with the conformal symmetry assumption \cite{Polyakov:1970xd}. 

In the scale invariant context, the skeleton expansion idea goes back to Patashinskii and Pokrovskii \cite{Patashinskii1964}, and it was later discussed by Polyakov \cite{Polyakov1968}, Migdal \cite{Migdal1968}, Gribov and Migdal \cite{Gribov:1968fg}. With just scale invariance the idea is not strong enough. There are too many parameters in the exact vertex, and one has to resort to ad hoc approximations. But with conformal invariance, Polyakov's paper \cite{Polyakov:1970xd} showed that the 3-point vertex is, up to an overall constant, exactly known in terms of the scaling dimensions of the fields, Eq.~\eqref{eq:3pt}. This was a big boost, and in 1971 Migdal \cite{Migdal:1971fof} and Parisi and Peliti \cite{Parisi:1971zza} proposed concrete computational schemes for critical exponents, followed in 1972 by further advances \cite{Symanzik:1972wj,Mack:1972kq,Parisi:1972zm}. The first lowest-order calculation of $O(2)$ critical exponents was performed with this method in \cite{Deramo1971}.
Migdal \cite{Migdal:1971fof} called this procedure ``bootstrap'', presumably in analogy with the S-matrix bootstrap of Geoffrey Chew. It is nowadays referred to as the ``old bootstrap'', to distinguish from the modern conformal bootstrap which we will come to in \cref{sec:birth}.

This spike of enthusiasm turned out to be short-lived. In 1971, Ken Wilson \cite{Wilson1971} ushered in a new understanding of critical phenomena through the renormalization group (RG). In just a few years, this became a dominant paradigm, not least thanks to the $\epsilon$-expansion of Wilson and Fisher \cite{Wilson:1971dc}, which endowed the RG with a small parameter. 

In his 1982 Nobel lecture \cite{Wilson1982}, Ken Wilson reserved some words of praise and some words of caution for the "old bootstrap": 

\begin{mybox} \small If the 1971 renormalization group ideas had not been developed, the Migdal-Polyakov bootstrap would have been the most promising framework of its time for trying to further understand critical phenomena. However, the renormalization group methods have proved both easier to use and more versatile, and the bootstrap receives very little attention today. $\langle\ldots\rangle$ the problem of convergence
	of the skeleton expansion leaves me unenthusiastic about pursuing the bootstrap
	approach, although its convergence has never actually been tested.---{\it Ken Wilson} (1982)
	\end{mybox}

It is actually possible to marry the ``old bootstrap'' with the $\epsilon$-expansion, done by Mack who considered the $\phi^3$ theory in $6+\epsilon$ dimensions \cite{Mack:1973kaa}. For recent work in this direction see \cite{Goncalves:2018nlv}. Alternatively, one can bring in a small parameter by considering a large $N$ limit. One of the most nontrivial applications of the ``old bootstrap'' is the computation of the $O(N)$ model $\eta$ exponent at order $1/N^3$ \cite{Vasilev1982}. See also \cite{Liendo:2021egi} for interesting recent work, a review of the ``old bootstrap'' and more references to papers using this technique. The interplay between $6-\epsilon$ and $1/N$ expansion was studied in Ref.~\cite{Fei:2014yja}.

\section{1974 - Polyakov's ``Non-Hamiltonian approach'' - the birth of modern conformal bootstrap}
\label{sec:birth}

The next landmark of our story is Polyakov's 1974 paper ``Non-hamiltonian approach to conformal quantum field theory'' \cite{Polyakov:1974gs}, whose short introduction reads as a manifesto:
\begin{mybox}
	\small
	In recent years, the hypothesis of the conformal invariance of strong interactions at distances much
	shorter than $10^{-14}$cm has been put forward and analyzed
	in detail. It has been shown that
	the equations of quantum field theory are invariant under
	the conformal group, under the condition that anomalous
	values of the dimensions, which should be determined
	from the condition for solubility of the equations, are
	assigned to the different fields. All the observable consequences of the theory were expressed in terms of these dimensions and, in addition, in terms of a set of
	effective interaction constants at short distances.
	
	\qquad At the same time, the equations for the determination
	of the above quantities (skeleton expansions for the
	vertex parts) were series with zero radius of convergence and therefore did not have well-defined mathematical meaning. The physical meaning of these equations was also highly obscure. The form of the equations depended in an essential way on the type of fundamental
	fields and on the form of their bare interaction, whereas
	the results of a theory with anomalous dimensions should
	not be sensitive to the choice of the initial Hamiltonian.
	
	\qquad The purpose of the present article is to construct a
	more general formalism for the determination of the
	anomalous dimensions; this, on the one hand, would be
	"democratic" with respect to the different fields, and,
	on the other, would not contain meaningless series
	(these two properties turn out to be intimately related).
	Compared with the old approach, such a formalism plays
	the same role as the methods of S-matrix theory compared with Hamiltonian theory, and is a generalization of the S-matrix equations for the short-distance region.---{\it A.M.~Polyakov} (1974)
\end{mybox}

So, Polyakov disavows the skeleton expansion approach, i.e.~"the old bootstrap." This stand (as well as the title of his paper) reminds of the oft-quoted passage from Landau's last published work "On the fundamental problems" (1960) \cite{Landau1960}:  
\begin{mybox} \small
	The Hamiltonian method for strong interactions has outlived itself and should be buried with, of course, all the honors it deserved.---{\it L.D.~Landau} (1960)
	\end{mybox} 

Uncharacteristically for Polyakov, the above introduction does not mention critical phenomena, focusing instead on the hypothetical connection to hadronic physics. This was to expire very shortly with Gross, Wilczek and Politzer's discovery of asymptotic freedom. But the new dynamical principle he's about to propose is completely general. 

Polyakov postulates that there is a set of local primary operators, which includes both scalars and operators with nonzero Lorentz spin. The primary operators and their derivatives form a complete set of local operators. This primary operator basis is orthogonal. He considers the operator product expansion (OPE) of primaries and points out that the coefficient functions of primaries ("C-functions") in the r.h.s. of the OPE are fixed by conformal invariance up to a few constants. He points out that the OPE allows to reduce four-point functions and other $n$-point functions to the three-point functions whose explicit form is known thanks to his 1970 work. He makes a very important observation that OPE is not asymptotic but has a finite radius of convergence. Finally, he states the dynamical equation on the OPE C-functions---that the four-point function possesses crossing symmetry after substituting into it the OPE for the different pairs of operators:

\begin{mybox}\small
	Finally, our program consists in calculating all the functions C to within a few constants, substituting the operator expansion into the four-point function and finding the unknown constants from the crossing-symmetry requirement.---{\it A.M.~Polyakov} (1974)
	\end{mybox}

The above credo is that of the modern conformal bootstrap, although Polyakov does not yet use this term---this terminology is due to \cite{Belavin:1984vu}. Nowadays, we would proceed to expand the four-point function into conformal blocks and impose the equality between the s- and t-channel expansions. In 1974, Polyakov's pursued a rather different implementation. The term ``conformal block'' does not appear in his work but he uses the terms	``algebraic amplitude'' and ``unitary amplitude''. He works in Lorentzian metric and uses dispersion relations to construct these basic building blocks for conformal correlators.  Algebraic amplitude is so called because it satisfies the ``algebra'', i.e.~the OPE---this corresponds to the modern conformal blocks. He notes that ``the algebraic amplitude possesses anomalous singularities in coordinate space,'' a downside for him as he wants an expansion converging everywhere in the Euclidean region. So he proceeds to construct ``unitary amplitudes'' which do not have anomalous singularities but which violate the OPE by logarithmic terms. He proposes (his Eq. (8.1))) to construct the full crossing symmetric amplitude by summing the s-, t- and u-channel unitary amplitudes and requiring that the logarithmic OPE-violating terms to cancel in the sum. His discussion is rather technical, and the final equations look forbidding. He gives however one nontrivial check of his formalism, reproducing the lowest-order $O(n)$ model $\epsilon$-expansion results of Wilson and Fisher. 

For the next 40 years, Polyakov's paper was a strong inspiration for the bootstrap philosophy, although not for the implementation details. Indeed, there was hardly any work on Polyakov's ``unitary amplitudes'' until when this line of thought was spectacularly revived in \cite{Sen:2015doa,Gopakumar:2016wkt,Gopakumar:2016cpb} under the name Polyakov-Mellin bootstrap. There, this was used to produce further $\epsilon$-expansion orders for Wilson-Fisher CFT scaling dimensions and OPE coefficients, going in some cases beyond RG predictions. Furthermore,  \cite{Mazac:2019shk,Caron-Huot:2020adz} defined ``Polyakov-Regge blocks'', which are closely related to Polyakov's unitary amplitudes.

\section{The Rome group}

In this and the next section we would like to mention the 1970s work on nonperturbative CFTs by other groups. The first group, in Rome, was led by Raoul Gatto (1930-2017) and included three young researchers: Sergio Ferrara (born 1945), Aurelio Grillo (1945-2017) and Giorgio Parisi (born 1948). They produced some 15 papers on CFT between 1971-75. Some of their main achievements were:
\begin{itemize}
	\item
	FGG 1971: Manifestly conformally invariant OPE \cite{Ferrara:1971vh}. Here they worked out in closed form, to all orders in the derivative expansion, the OPE coefficients for a spin-$\ell$ primary appearing in the OPE of two scalar primaries. They give two methods - by using the conformal algebra and from consistency with the three-point function.\footnote{Surprisingly, they did not cite the foundational Polyakov's 1970 paper \cite{Polyakov:1970xd}, nor Migdal \cite{Migdal:1971xh} who was the first to note the possibility of extracting the OPE from the 3-point function. According to Migdal's bitter memories \cite{Migdal-Gatto}, Gatto was the editor of his paper, and Ferrara and Grillo the referees.} 
	\item FP 1972: The shadow formalism \cite{Ferrara:1972xe} (see also \cite{Ferrara:1972uq}).
	\item FGGP 1972: Covariant expansion for the four-point function \cite{Ferrara:1972kab}. This is the first work to  discuss the modern conformal blocks (called by them ``conformal partial waves''), computed using the shadow formalism. This line of study continued in FGGP 1974 \cite{Ferrara:1974nf} and FGG 1975 \cite{Ferrara:1974ny}, with several explicit formulas, such as conformal blocks for a scalar exchange, conformal blocks in $d=2$, etc.
	\item
	FGG 1974: Positivity constraints on the anomalous dimensions \cite{Ferrara:1974pt}, deriving for the first time the unitarity bound $\Delta\ge \ell+2$ (in 4 dimensions).
	\end{itemize}
	
Several times, the Rome group noticed the crossing relation constraint for the four-point function:
	\begin{mybox}\small
	The four-point correlation function has to satisfy an additional crossing constraint besides those imposed by the space-time (and other) symmetries. \cite{Ferrara:1972kab}
		\end{mybox}
		\begin{mybox}\small
		One needs a complete discussion of the so-called "crossing relations" resulting on the four-point function when its four local operators are in various ways associated in pairs, Wilson-expanded, and the different outcomes compared. \cite{Ferrara:1973eg}
			\end{mybox}
		\begin{mybox}\small
		For a $n$-point function the causality restrictions of the theory imply sets of equalities, resembling crossing relations, to be satisfied by the coefficients of the irreducible representations which contribute to the expansions. \cite{Ferrara:1973yt}
	\end{mybox}
	
And finally, in what was to be their last paper on the subject:
\begin{mybox}\small
Polyakov has proposed different choices of partial waves, of different analyticity properties, and obtained dynamical constraints (self-consistency conditions) in conformal invariant theories. Similar dynamical constraints have also been suggested by us. The aim of the present paper is to further investigate the singularity structure of conformal partial-wave amplitudes. 

\qquad We find that, for any value of the dimension of space-time $D > 2$, Euclidean singularities are present in partial-wave amplitudes, this fact reflecting the lack of convergence of conformal operator expansion at large distances.

\qquad Our partial-wave amplitudes, to be considered as related to Wightman functions in Minkowski space, are unambiguously defined using the conformal ansatz for the operator expansion, regarded as a Taylor expansion near the tip of the light-cone, which verifies the Wilson dimensional rule. We stress that these requirements avoid confusion with different possible ansatz for partial-wave amplitudes. Because of the exhibited singularity structure, which reflects itself into a violation of causality in single partial waves, it is clear that dynamics must provide a mechanism of cancellation of such singularities. Whether such a problem has any nontrivial solution is still at present an open question. Free-field theories provide at least an example. \cite{Ferrara:1974ny}
\end{mybox}

So who invented the modern conformal bootstrap? 

Polyakov was very forceful in expressing the bootstrap philosophy/program. Working in the expansion basis of unitary blocks he did obtain an example where his equations worked. While it took time, his ideas were fully vindicated and became a branch of modern bootstrap (the Polyakov-Mellin bootstrap). 

Gatto's obituary \cite{maiani_gatto_2024} by Luciano Maiani says that the Rome group also ``formulated the Conformal Bootstrap program.'' I am not sure about the full strength of this statement. The Rome group did say that crossing needs to be satisfied in a consistent theory, but it's not clear how they felt about raising it to the status of a dynamical principle. They did not have any example of solving crossing, leading to nontrivial anomalous dimensions. Giorgio Parisi recalls that they were ``stuck'' \cite{Parisi:2018kqi} at finding an implementation. To their credit, they did emphasize using the conformal blocks---the most frequently used basis nowadays.  But cancellation of conformal block singularities which they mentioned in \cite{Ferrara:1974ny} does not look promising, and as of today nobody managed to make it work. We will see in \cref{sec:numCB} the alternative strategy which proved successful.

\section{Mack and the Sofia group}\label{sec:mack-and-the-sofia-group}
Gerhard Mack was among the early CFT proponents. In 1969 he and Abdus Salam found the infinitesimal transformation rules of primary fields, derived from the representation theory point of view \cite{Mack:1969rr}, textbook material nowadays \cite{DiFrancesco:1997nk}. He then contributed to the development of the ``old bootstrap'' \cite{Mack:1972kq,Mack:1973kaa}, \cref{sec:old-bootstrap}, and wrote an early review of this approach \cite{Mack-1973-old-bootstrap-review}. Several of his other incisive contributions to CFT from the 1970s are:

\begin{itemize}
	\item A rigorous classification of all unitary conformal representations in $d=4$ \cite{Mack:1975je}. This classic paper is usually cited for the unitarity bounds, along with Minwalla \cite{Minwalla:1997ka} who generalized to other $d$ and to the superconformal algebra (Minwalla's paper only proves the necessary part of the necessary and sufficient conditions for unitarity). For completeness, the early work by the Rome group \cite{Ferrara:1974pt} should be also mentioned. In mathematics literature, results equivalent to general $d$ unitarity bounds were derived by Jantzen \cite{jantzen_kontravariante_1977}, although the relevance of this work for physics was realized only recently \cite{Penedones:2015aga,Yamazaki:2016vqi}.
	
	\item In another classic work \cite{Luscher:1974ez}, with his PhD student Martin L\"uscher (born 1949), Mack resolved a long-standing puzzle about global properties of conformal group in Lorentzian signature. They showed that the appropriate arena for Lorentzian CFT in $d$ dimensions is the Lorentzian cylinder $S^{d-1}\times \mathbb{R}$ of which the (conformally compactified) Minkowski space is but a patch. 
	
	\item In \cite{Mack:1976pa}, Mack proved the strong convergence of OPE in CFTs, assuming Wightman axioms in the Lorentzian, conformal invariance in the Euclidean, and an OPE in asymptotic sense. The proof may have a gap \cite[Sec.~8.3]{Kravchuk:2021kwe}.
	\end{itemize}

Mack developed a yet another expansion for the conformal four-point functions, into conformal partial waves associated with the unitary representations of the Euclidean conformal group \cite{Mack1974}. The principal series of unitary representations being parametrized by a continuous range of scaling dimensions $\Delta = d/2+i\mathbb{R}$, Mack's new expansion takes the form of an integral over this continuous range which extends in the imaginary $\Delta$ directions. This representation is nowadays called the ``Euclidean inversion formula'' \cite{Caron-Huot:2017vep}. Mack hoped (and contemporary researchers concur, although this has not been proven yet) that the integration amplitude is meromorphic with a discrete series of poles on the real $\Delta$ axis, and appropriate asymptotics at infinity, so that by deforming and closing the contour along the real axis, one could make contact with the usual OPE in the CFT. Mack's approach became known as the ``harmonic analysis on the conformal group'', from the title of the book he coauthored with a group of colleagues from Sofia led by Ivan Todorov (1933-2025) and including Vladimir Dobrev and Valentina Petkova \cite{Dobrev:1977qv}. The book is a mix of the ``old bootstrap'' with nonperturbative ideas. In \cite[Sec.~16]{Dobrev:1977qv}, they write a nonperturbative equation for the crossing symmetry for the four-point function:
\beq
\myinclude{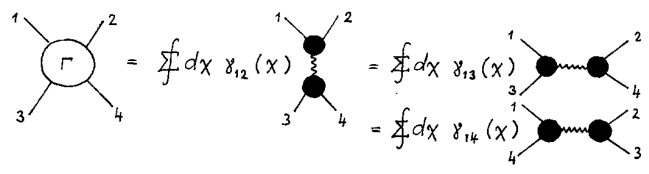}\,.
\eeq
Their discussion, based on the paper \cite{Dobrev:1975ru} by the Sofia group, is quite modern. E.g.~they introduce the crossing kernel for the conformal partial waves. Unfortunately, this was not pursued further at the time.\footnote{In 2009, at the age of 69, Mack made an impressive comeback to CFT with the work on Mellin amplitude representation of conformal correlators \cite{Mack:2009mi,Mack:2009gy}. This proved extremely useful, in particular in connection with the  AdS/CFT correspondence \cite{Penedones:2010ue}. See \cite{Penedones:2019tng} for recent work in this direction.}

\section{1984 - Belavin, Polyakov, Zamolodchikov}

In the second half of the 1970s and the beginning of the 1980s, some of the main CFT players turned their attention elsewhere: Polyakov to gauge theories and strings, Ferrara to supersymmetry and supergravity, Parisi to spin glasses.

This hiatus was followed by an explosion in 1984, when Belavin, Polyakov and Zamolodchikov (BPZ) published ``Infinite Conformal Symmetry in Two-Dimensional Quantum Field Theory'' \cite{Belavin:1984vu}, one of the most beautiful and influential papers in theoretical physics of the 20th century.

This classic paper is textbook material \cite{DiFrancesco:1997nk}, so I will give a very brief summary. The symmetry algebra of the 2D CFT is infinite-dimensional Virasoro algebra with a central charge $c$, which extends the finite-dimensional global conformal algebra. The new concept is a Virasoro primary field. The Virasoro algebra multiplet built on top of a Virasoro algebra contains infinitely many global conformal algebra multiplets. They introduce Virasoro conformal blocks corresponding to exchanges of the Virasoro primary and all of its Virasoro descendants. These conformal blocks can be expanded order-by-order in a short-distance expansion, with coefficients fixed by Virasoro algebra. They write the crossing equation relating the s- and t-channels and call it ``the bootstrap''. They introduce "degenerate" Virasoro primaries---primaries of special dimensions (depending on $c$) for which the multiplet contains less states. This leads to correlators of degenerate primaries satisfying differential equations. They notice that degenerate primaries close under OPE and define ``minimal theories'' where all primaries are degenerate. This happens for special discrete sequence of the central charge $c$ between 0 and 1. The simplest minimal model of $c=1/2$ is identified with the critical 2D Ising model CFT, having three Virasoro primaries: $\mathds{1}$, $\epsilon$ and $\sigma$. Their scaling dimensions are fixed by the degeneracy assumption at $0,1,1/8$, and only OPE coefficients are left to be determined. The important OPE is
\beq
\sigma\times\sigma=\mathds{1}+\lambda_{\sigma\sigma\epsilon} \epsilon\,.
\eeq
 In an appendix they solve the conformal bootstrap equation in this case. They compute in closed form the Virasoro blocks $G_{\mathds{1}}$ and $G_{\epsilon}$ for the s-channel expansion of the four-point function of $\sigma$, schematically:
 \beq
 \langle \sigma\sigma\sigma\sigma\rangle = G_{\mathds{1}}+ \lambda^2_{\sigma\sigma\epsilon} G_{\epsilon},
 \eeq
 and find their linear combination which is crossing-invariant, which sets $\lambda_{\sigma\sigma\epsilon}=1/2$.
 
 The minimal model assumption of BPZ was rationalized in the subsequent work by Friedan, Qiu, and Shenker \cite{Friedan:1983xq}, who showed that unitary 2d CFTs with central charge $c<1$ have to be minimal models and have central charge $c=1-\frac{6}{m(m+1)}$, $m=3,4,5,\ldots$, with $m=3$ corresponding to the 2D Ising model.\footnote{The constraint that $c\ge 1/2$ in unitary 2d CFT was also obtained in 1976 in an unpublished work by L\"uscher and Mack \cite{luscher_mack_1976}.}
 
 The minimal models and, more generally, rational CFTs, provide an amazing class of theories where crossing can be solved with finitely many primary fields. The literature about them is huge \cite{DiFrancesco:1997nk} and keeps growing. However, many researchers believe that rational CFTs are special, and that there are many interesting $c>1$ CFTs which are irrational. For a simple yet nontrivial example, couple three critical 3-state Potts models via the relevant operator $\epsilon_1\epsilon_2+\epsilon_1\epsilon_3+\epsilon_2\epsilon_3$ and flow to the IR fixed point. Perturbation theory and lattice simulations \cite{Dotsenko:1998gyp} estimate the IR central charge $c_{\rm IR}\approx 2.38$, only slightly below $c_{\rm UV}=4/5\times 3=2.4$. It seems likely that this IR CFT is irrational, although this has not been proven yet.\footnote{See \cite{Antunes:2022vtb,Antunes:2025huk} for related recent work.}
 
 Can we do conformal bootstrap in $d=2$ without the rationality assumption, i.e.~for infinitely many Virasoro primaries? And can we do conformal bootstrap in $d>2$, where only the global conformal algebra is available, and the number of global conformal primaries is always infinite? Concerning the latter, BPZ left a warning: 
 
 \begin{mybox}In the multidimensional case $d>2$, the system proves to be too complicated to solve exactly, the main difficulty being the classification of the fields entering the algebra. \cite{Belavin:1984vu}
 	\end{mybox}

\section{1980s and 1990s developments in CFTs in $d>2$ dimensions}

In 1980s and 1990s, there were many interesting results in CFTs in $d>2$ dimensions, although no attempt to bootstrap them was done at the time. The partial list includes:
\begin{itemize}
	\item Banks and Zaks \cite{Banks:1981nn} considered conformal fixed points of Yang-Mills theories in 4d with a number of matter fields close to the upper bound allowed by asymptotic freedom, known today as the Banks-Zaks fixed points. The same idea was expressed in 1974 by Caswell \cite{Caswell:1974gg} and by Belavin and Migdal \cite{Belavin:1974gu}.
	
	\item It was shown that the maximally supersymmetric 4d Yang-Mills theory ($\mathcal{N}=4$ super Yang-Mills) is conformal (see \cite{Brink-story}). This is interesting because the theory has an exactly marginal coupling constant. We thus have a manifold of conformal fixed points. 
	
	\item Seiberg \cite{Seiberg:1994pq} studied IR phases of $\mathcal{N}=1$ super-QCD theories in 4d and discovered IR dualities, i.e.~when two different UV theories flow to the same IR fixed point.
	
	\item Maldacena \cite{Maldacena:1997re} proposed a duality between large-N CFTs and theories of gravity in AdS (AdS/CFT correspondence). This raised enormously the interest and awareness about CFTs in $d>2$ dimensions, even though this has not had immediate influence on the development of the bootstrap.\footnote{Later on AdS/CFT and conformal bootstrap had fruitful interactions. An early example is \cite{Heemskerk:2009pn}. Some AdS/CFT predictions were shown to be generally true in appropriate asymptotic limits \cite{Fitzpatrick:2012yx,Komargodski:2012ek,Pal:2022vqc,vanRees:2024xkb}. Also, it was understood \cite{Gopakumar:2016wkt,Gopakumar:2016cpb,Gopakumar:2018xqi} that Polyakov's unitary blocks from \cite{Polyakov:1974gs} are basically AdS Witten diagrams \cite{Witten:1998qj} corrected by contact terms. On the other hand, the usual conformal blocks are computable by ``geodesic Witten diagrams'' \cite{Hijano:2015zsa}.}
	
	\end{itemize}

\section{2006 - Conformal technicolor and the numerical conformal bootstrap}
\label{sec:numCB}

\subsection{Personal background}

In 2002 I obtained my PhD in mathematics from Princeton University, on a problem in harmonic analysis which I solved during my first year. My official advisor was Elias Stein, but from the second year on I worked in theoretical physics with Sasha Polyakov who became my unofficial advisor. We worked on the gauge theory loop equation in AdS/CFT, and some aspects of string theory. After the PhD I continued in physics, with the first postdoc in the Amsterdam string theory group. I did some more string theory work which got me an invitation to talk at Strings 2005 in Toronto \cite{Grant:2005qc}. But overall I was not very satisfied with strings, as I wanted a more immediate connection with experiments. I started to work on possible TeV-scale black hole production at the upcoming LHC \cite{Rychkov:2004sf,Giddings:2004xy,Yoshino:2005hi}. This may sound like crazy wishful thinking nowadays, but at the time there were thousands of papers on TeV-scale gravity scenarios \cite{Arkani-Hamed:1998jmv}. Some joked that observation of mini black holes decaying at the LHC is the best chance for Stephen Hawking winning a Nobel prize \cite{Giddings:2002av}.

In 2005, I moved to a second postdoc at the Scuola Normale Superiore in Pisa, in the group of Riccardo Barbieri. His interests were in the electroweak phenomenology and Higgs physics. It was not immediately clear why I got hired,\footnote{I learned later that this was thanks to Massimo Porrati, who was there on a sabbatical and put in a word for me since he knew my black hole work.} but it was just what I needed. Intellectually stimulating theory and lots of data to cope with: old data from LEP, Tevatron, dark matter detection experiments, and the upcoming LHC. It was an exciting time for the electroweak pheno. People were trying to imagine what LHC could see once it starts, beyond plain vanilla Higgs boson or minimal SUSY. 
Randall-Sundrum, large extra dimensions, composite Higgs, little Higgs, Higgsless models, you name it\ldots. The "LHC Olympics" were organized by Nima Arkani-Hamed to prepare the US phenomenologists to deciphering the LHC signal (string theorists also participated). The European theorists were also preparing, although not at such a flamboyant level. 

I soon got to know many interesting people in the hep-ph community. One of the most remarkable new acquaintances was Riccardo Rattazzi. We first met at a couple of meetings in late 2005/early 2006, and he invited me to visit him at CERN, to work on ``Conformal Technicolor.'' Little did we suspect that this project would lead to the revival of the conformal bootstrap.

\subsection{Conformal Technicolor}
Conformal Technicolor, an idea for beyond the Standard Model (BSM) physics, was proposed in 2004 by Markus Luty and Takemichi Okui \cite{Luty:2004ye}. To explain it, consider the Higgs field mass term and the Yukawa interactions (written schematically) in the SM Lagrangian:
\beq
\mathcal{L}_{SM}\supset m^2|H|^2+y \psi_L H \psi_R\,.\label{eq:SM}
\eeq
In the SM, the Higgs field $H$ has classical scaling dimension 1, and consequently $|H|^2$ has scaling dimension 2. That $\Delta_H=1$ is great for the success of SM Yukawa interactions describing flavor physics, in particular the experimental limits on flavor changing neutral currents (FCNC). On the other hand, that $|H|^2$ has scaling dimension 2 means that the Higgs mass term is relevant, which is the origin of the naturalness problem of the SM.

The Higgs boson was discovered at the LHC in 2012, and its properties were measured to be in agreement with the SM. So, the SM is by now confirmed, at least to a cutoff of a few TeV, and its naturalness problem is a fact of life. But back in 2006 most hep-ph researchers were still hoping that the naturalness problem will be somehow resolved at a scale accessible to the LHC. We were thinking about possible mechanisms of its resolution, in the hope of guessing the correct BSM model.

Consider for instance the Technicolor scenario, in which the Higgs field is not fundamental but is a techni-fermion bilinear, $H=T\bar T$, of scaling dimension $3$. Then the Higgs mass term is irrelevant, and the naturalness problem is absent. However, we get a problem with the Yukawa couplings which are now four-fermi operators of dimension 6:
\beq
\frac{y}{\Lambda_{\rm F}^2}\psi_L T \bar T \psi_R\,,
\eeq
where $\Lambda_{\rm F}$ is a new "flavor" scale, which cannot be too high, since there are some heavy fermions in the SM model (the top quark). One expects that there will be FCNC effects associated with this scale, and this leads to a tension with the absence of such effects in experiments. That's one of the reasons why Technicolor was disfavored long before the LHC started colliding protons. (Another reason is that generically it gives a too large contribution to the so-called S-parameter in the electroweak precision tests.)

The Luty-Okui idea was to aim for a viable model in between the SM and the Technicolor.\footnote{Their idea can also be viewed as an abstraction and a relaxation of an earlier Walking Technicolor, which in interest of time we do not discuss here.} They replace the Higgs sector of the SM with a strongly interacting CFT which has an $SO(4)=SU(2)\times SU(2)$ global symmetry,\footnote{This is a ``custodial'' $SO(4)$ which also the SM Higgs sector has.} and a primary operator $H$ which transforms as a vector of $SO(4)$. The scaling dimension of $H$ is assumed to be \beq
\Delta_{H}=1+1/\text{few}.\label{eq:H}
\eeq 
In a strongly coupled theory, we need to explain what we mean by the operator $|H|^2$. We define it as the lowest-dimension operator, after the unit operator, which appears in the OPE $H\times H^\dagger$ and which is a singlet of $SO(4)$:
\beq
H\times H^\dagger \supset \mathds{1}+|H|^2+\ldots
\eeq
Luty and Okui assume that
\beq
\Delta_{|H|^2}\gtrsim 4.\label{eq:H2}
\eeq 
Note that combined with \eqref{eq:H}, this needs a significant deviation from the relation $\Delta_{|H|^2}=2 \Delta_{H}$. This would be impossible at weak coupling or at large N. But Luty and Okui hypothesized that a strongly coupled and small-N CFT might exist (the latter is also needed to mitigate the S-parameter), in which both \eqref{eq:H} and \eqref{eq:H2} hold. Then, taking such a CFT, coupling it to the SM via $SU(2)\times U(1)$ gauge interactions, and turning on the Yukawa and Higgs mass term as in \eqref{eq:SM}, would provide a beautiful theory of electroweak symmetry breaking. This theory does not suffer from the naturalness problem, because of \eqref{eq:H2}. Nor is it killed by flavor constraints, since due to \eqref{eq:H} the Yukawa term is only weakly irrelevant, and the flavor scale $\Lambda_{\rm F}$ can be much higher than in Technicolor, pushing FCNC effects below experimental limits. 

Brilliant! But do such theories exist? As Luty-Okui noticed \cite{Luty:2004ye} :
\begin{mybox}
	We are therefore led to a rather dark corner of theory space: non-supersymmetric 4D strongly-coupled conformal field theories with small N. $\langle\ldots\rangle$ Not much is known about the dynamics of such theories, and so our discussion of these theories is necessarily speculative.
\end{mybox}

\subsection{Numerical conformal bootstrap}

In August 2006 at CERN, Riccardo Rattazzi and myself started discussing if there was any way to probe this scenario. If $\Delta_H=1$, the theory is free and so $\Delta_{|H|^2}=2$. Can we somehow exhibit continuity in the limit $\Delta_H\to1$? For example, is there any upper bound on how much $\Delta_{H^2}$ may deviate from 2 when $\Delta_H$ starts deviating from 1? For simplicity we postponed the $SO(4)$ global symmetry case, replacing it by $\mathbb{Z}_2$, with $H$ a real scalar.  Rather quickly we realized that it's a conformal bootstrap problem. Neither of us was a CFT expert, but we knew the basics. Riccardo worked through the BPZ paper in 1985 for his 3rd year research project as an undergraduate at the Scuola Normale. Former Polyakov's student, I was among the lucky few who knew about his ``Non-hamiltonian'' paper \cite{Polyakov:1974gs}.\footnote{In 2006, the paper had 21 citations on INSPIRE-HEP, it now has over 500.}

Consider the Euclidean CFT four-point function:
\beq
\langle H(x_1) H(x_2) H(x_3) H(x_4)\rangle = \frac{1}{x_{12}^{2\Delta_H} x_{34}^{2\Delta_H} }g(u,v)\,,
	\eeq
	where $u=\frac{x_{12}^2 x_{34}^2}{x_{13}^2 x_{24}^2}$ and $v=u|_{1\leftrightarrow 3}$ are the conformal cross ratios. The function $g(u,v)$ can be expanded in conformal blocks as:
	\beq
	g(u,v)=1+\sum_{\Delta,\ell} \lambda_{\Delta,\ell}^2 g_{\Delta,\ell}(u,v)\,,
	\eeq
where 1 is the unit operator contribution, $\Delta>0,\ell=0,2,4,\ldots $ are the dimensions and spins of all nontrivial operators appearing in the OPE $H\times H$, $\lambda_{\Delta,\ell}$ are their OPE coefficients which are real numbers. This is the so-called s-channel conformal block expansion, which corresponds to the OPE $H(x_2)\times H(x_1)$ and converges at least in the domain where 
\beq
|x_2-x_1|<\min (|x_3-x_1|,|x_4-x_1|). \label{eq:schannel}
\eeq 
(in fact in a much larger domain but this will not be important for the present discussion). Crossing constraint means that 
\beq
\frac{1}{x_{12}^{2\Delta_H} x_{34}^{2\Delta_H} }g(u,v) = \frac{1}{x_{23}^{2\Delta_H} x_{14}^{2\Delta_H} }g(v,u)\,,
\eeq
where the r.h.s.~of this equation can be evaluated most naturally using the OPE $H(x_2)\times H(x_3)$ and can be expressed by the ``t-channel'' conformal block expansion which converges in the domain including
\beq
|x_2-x_3|<\min (|x_2-x_1|,|x_4-x_1|) \label{eq:tchannel}
\eeq 
Importantly, the convergence domains \eqref{eq:schannel} and \eqref{eq:tchannel} overlap.

Equating the two conformal block expansion we get the bootstrap constraint 
\beq
\label{eq:bootstrap}
v^{\Delta_H}\left[1+\sum_{\Delta,\ell} \lambda_{\Delta,\ell}^2 g_{\Delta,\ell}(u,v) \right]= u^{\Delta_H}\left[1+\sum_{\Delta,\ell} \lambda_{\Delta,\ell}^2 g_{\Delta,\ell}(v,u)\right]\,.
\eeq
Note that since the equation depends on $\Delta_H$, it gives hope to learn something about the spectrum of operators in the OPE $H\times H$, and in particular about the operator $H^2$ defined as the lowest dimension scalar $(\ell=0)$ in this OPE. 

While we quickly got to this point, to make further progress we needed the 4D conformal blocks! The 2D global conformal blocks were worked out by Ferrara, Gatto and Grillo in 1975 \cite{Ferrara:1974ny}, who expressed them in terms of hypergeometric ${}_2F_1$ functions of the lightcone coordinates, in a factorized left-moving$\times$right-moving form, naturally so because the 2D conformal group factorizes. As we were delighted to find out after a couple of internet searches, the 4D blocks were also worked shortly before we needed them, in 2001 by Francis Dolan and Hugh Osborn\footnote{Other prescient CFT works from the 90's by Hugh Osborn and his PhD students include \cite{Osborn:1993cr,McAvity:1995zd,Erdmenger:1996yc}.} \cite{DO1,DO2}, who found that they almost-factorize in the same lightcone coordinates, namely:
\begin{gather}
g_{\Delta,\ell}(u,v)=\frac{z \bar z}{z-\bar z}[k_{\Delta+\ell}(z) k_{\Delta-\ell-2}(\bar z)-(z\leftrightarrow \bar z)]\,,\\
u=z\bar z,\quad v=(1-z)(1-\bar z)\,,\label{eq:uv}\\
k_{\beta}(z)=z^{\beta/2} {}_2F_1(\beta/2,\beta/2,\beta,z)\,.
\end{gather}
This was extremely fortunate since we could now proceed without delay to the most intriguing part---the analysis of the bootstrap constraint \eqref{eq:bootstrap}.

We rewrote \eqref{eq:bootstrap} as a sum rule (recall \eqref{eq:uv}):
\begin{gather}
\label{eq:sumrule}
1= \sum p_{\Delta,\ell} F_{\Delta_H,\Delta,\ell}(z,\bar z),
\\
p_{\Delta,\ell} = \lambda_{\Delta,\ell}^2\ge 0\\
F_{\Delta_H,\Delta,\ell}(z,\bar z) = \frac{v^{\Delta_H} g_{\Delta,\ell}(u,v)-u^{\Delta_H} g_{\Delta,\ell}(v,u)}{u^{\Delta_H}-v^{\Delta_H}}\,.
\end{gather}
Then, plotting $F_{\Delta_H,\Delta,\ell}(z,\bar z)$ on the interval $0<z=\bar z<1$, for $\Delta_H$ close to 1, we noticed some interesting things:
\begin{itemize}
	\item
		$f_{\Delta,\ell}(z):=F_{\Delta_H,\Delta,\ell}(z,z)$ is symmetric around $z=1/2$. (By its definition.)
		\item
		$f_{\Delta,\ell}(1/2)>0$\,.
		\item
		$f_{\Delta,\ell}''(1/2)>0$ for $\ell=2,4,6,\ldots$ and $\Delta$ above the unitarity bound.
		\item
		 $f_{\Delta,\ell}''(1/2)<0$ for $\ell=0$ and $\Delta$ below some $\Delta_*$ which depends on $\Delta_H$, while for larger $\Delta$ it also become positive. Numerically, $\Delta_*\approx 3.6$.
\end{itemize}
From here, we concluded that, for $\Delta_H$ close to 1, any 4D CFT must have at least one scalar operator in the $H\times H$ OPE with $\Delta<\Delta_*$. Indeed, all $f_{\Delta,\ell}(z)$ should sum up, with positive coefficients $p_{\Delta,\ell}$, to a function identically equal to 1, which therefore has zero second derivative. But they wouldn't be able to do so if all terms in the sum had positive second derivative!

This result, obtained by the end of my August 2006 stay at CERN, looked quite encouraging. While the numerical value of $\Delta_*\approx 3.6$ was weaker then the expectation that the upper bound on $\Delta_{H^2}$ should approach 2 as $\Delta_H\to 1$, we obtained it by using only very partial information about $F_{\Delta_H,\Delta,\ell}$. We had all reasons to hope that more detailed information would lead to stronger constraints. Still at CERN, we realized that the sum rule in presence of the positivity constraint $p_{\Delta,\ell}\ge 0$ is a Linear Programming problem which, if properly discretized, could be solved numerically via the Simplex Method.\footnote{There was a chapter about Linear Programming in the book \cite{strang1976linear}, whose Russian translation was a favorite Linear Algebra text of my father, an aircraft engineer. I must have picked it up from there.} A dual formulation---to search for a linear functional non-negative on all conformal blocks---looked particularly convenient from the point of view of establishing rigorous bounds.

\begin{figure}
	\centering
	\includegraphics{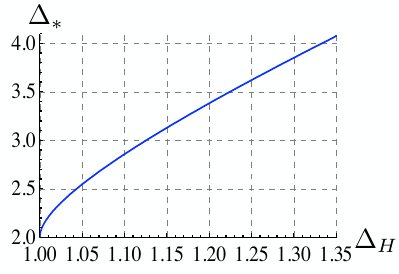}
	\caption{The bound from \cite{Rattazzi:2008pe}. Figure adapted from \cite{Rychkov:2009ij} where this bound was further improved.}
	\label{fig:first-bound}
	\end{figure}

While the program was clear, its completion took almost two years. Erik Tonni, my officemate in Pisa, joined the project in November 2006, and we started to improve the bounds. For about a year, Erik and I played with imposing the sum rule at a collection of points scattered around $z=\bar z=1/2$. We got somewhat better results with several points along $z=\bar z$, and further improvements by adding points at $z\ne\bar z$. However, the numerics were not very stable, and it was not clear how to distribute the points to get a systematic improvement. In December 2007 we finally realized what in retrospect looks like an obvious generalization of the CERN argument---that we should simply impose the sum rule in the Taylor expansion around $z=\bar z=1/2$, up to some finite order in both $z$ and $\bar z$. By increasing the Taylor expansion expansion the bounds were guaranteed to get stronger and stronger---a clear advantage. The subsequent progress was rapid and we got much improved bounds, which were now clearly approaching $2$ as $\Delta_H\to 1$. In February 2008, Erik and I visited Riccardo at the EPFL to discuss these results. We also invited Alessandro Vichi, a first-year PhD student supervised by Riccardo, to join the project. In the ensuing months, Alessandro contributed crucially to strengthen the bounds and to streamline and double-check all the logic and the numerical machinery. The first draft of the paper was put together in March-April 2008 while I was visiting Stefan Pokorski's group in Warsaw, where I also presented this work in an informal seminar. In May 2008, we added the constraint from the large $\Delta$, $\ell$ asymptotics of the conformal blocks, worked out by Erik, which further stabilized the numerics. Finally, we felt ready to go public with our findings. This honor fell to Riccardo, who presented our work in a plenary talk at the Planck conference in Barcelona on May 23, 2008, while the paper was submitted to arXiv a month later \cite{Rattazzi:2008pe}. Our main result was the bound 
\beq
\Delta_{H^2}\le \Delta_*(\Delta_H),
\eeq
valid in any unitary 4D CFT, where the function $\Delta_*(\Delta_H)$ was computed numerically in the range $1\le \Delta_H\le 1.35$, see \cref{fig:first-bound}. 

The method we developed (expand the four-point function in conformal blocks, impose the crossing constraint up to some finite order in the Taylor series expansion around the point $z=\bar z=1/2$, and search numerically for non-negative linear functionals) became known as the numerical conformal bootstrap. Compared to previous work, the main differences were:
\begin{itemize}
	
	\item
	In the 2D minimal model case considered by BPZ \cite{Belavin:1984vu}, the scaling dimensions were exactly known, and crossing involved
	finitely many Virasoro blocks. The conformal bootstrap problem was then reduced to finding finitely many OPE coefficients, which could be solved exactly. This strategy was not available in higher dimensions, where the number of conformal primaries is always infinite.
	
	\item
Polyakov \cite{Polyakov:1974gs} and FGG \cite{Ferrara:1974ny} were concerned about the ``unphysical'' singularities of individual conformal blocks. (These singularities are visible in \eqref{eq:uv} as cuts $z,\bar z \in (1,\infty)$ of the ${}_2 F_1$ hypergeometric functions.) This led Polyakov to eschew the conformal blocks in favor of the unitary ones. Ref.~\cite{Ferrara:1974ny} mused about cancellation of these singularities once summed over the infinitely many blocks, but this was unworkable since the sum does not absolutely converge in this region. In our work, we did not chase these singularities at all, focusing instead on the region around $z=\bar z=1/2$ where both s- and t- channels converge (exponentially fast, as explained in later work \cite{Pappadopulo:2012jk}).

	\item 
  Instead of finding solutions to crossing, we were content to get bounds ruling out parts of the CFT parameter space. This was a small but concrete result, showing that the conformal bootstrap has nontrivial constraining power in $d>2$. (It was recognized later that the numerical conformal bootstrap does allow to recover approximate solutions to crossing by going to the boundary of the allowed region \cite{Poland:2010wg,El-Showk:2012vjm}.)
   
   \item 
  Our result was numerical (although we had analytic understanding of the $\propto \sqrt{\Delta_H-1}$ behavior of the bound as $\Delta_H\to 1$). However the amount of involved computations was modest. The success was mainly due to the realization that one has to work around $z=\bar z=1/2$, not to the progress in computing power. (A non-rigorous version of conformal bootstrap which is even more lightweight in computations was later proposed by Gliozzi \cite{Gliozzi:2013ysa}.)
	
	\end{itemize}

In September 2008 I gave a talk about our work at the IAS. Since no major objections were raised,\footnote{I recall Nati Seiberg was especially supportive. Polyakov had questions about the OPE convergence in presence of infinitely many exchanged primaries, but to me the argument via cutting and gluing along the spheres \cite{Polchinski:1998rq} seemed pretty robust. Later on, this was further elaborated in \cite{Pappadopulo:2012jk}.} we submitted the paper to JHEP, and it was published after a minor altercation with the Editor.

%
%
%
%
%
%
%
%

\section{2011-2014 - Attack on the 3D Ising Model}\label{sec:2011-2014---attack-on-the-3d-ising-model}

In the couple of years after \cite{Rattazzi:2008pe}, our group continued to pursue the numerical conformal bootstrap in $d=4$, consolidating the original findings and extending them in various new directions \cite{Rychkov:2009ij,Caracciolo:2009bx,Rattazzi:2010yc,Rattazzi:2010gj}. In 2010, David Poland and David Simmons-Duffin \cite{Poland:2010wg} became the first other group to use the new method, as well as extend it to superconformal theories in $d=4$. 
	David Poland visited me at the Ecole Normale Superieure in Paris in November 2010. In April 2011, the first "Back to the Bootstrap" workshop was organized at the Perimeter Institute by Jo\~ao Penedones, Leonardo Rastelli and Pedro Vieira. Bootstrap was gaining traction.

\begin{figure}[h]
	\centering
	\includegraphics[width=0.45\textwidth]{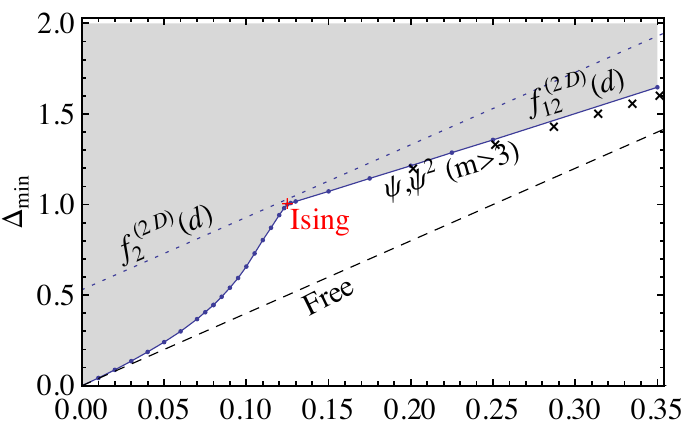}
	\caption{The $d=2$ bound from \cite{Rychkov:2009ij}, with a kink at the 2D Ising CFT.}
	\label{fig:2d-bound}
\end{figure}

\begin{figure}[h]
	\centering
	\includegraphics[width=0.45\textwidth]{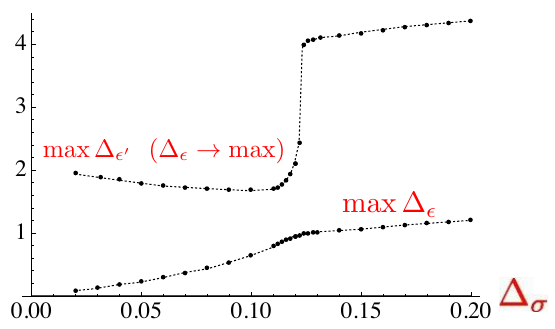}
	\caption{The bound on the second scalar ($\epsilon'$) with the first one ($\epsilon$) at the gap \cite{Rychkov:2011et}.}
	\label{fig:second-op-bound}
\end{figure}

In August 2011, I was invited to the "Scalars 2011" conference at the University of Warsaw, where most talks were about the Higgs boson and other hypothetical scalar \emph{particles}. Instead I decided to speak about scalar \emph{operators} in CFTs. With Alessandro Vichi we found back in 2009 \cite{Rychkov:2009ij} that in $d=2$ the bootstrap bound analogous to Fig.~\ref{fig:first-bound} showed a "kink"\footnote{Referred to at the time as a "knee".} at the position of the 2D Ising CFT, Fig.~\ref{fig:2d-bound}. My talk offered an explanation. Fixing the first exchanged scalar to the bound, I computed the bound on the second scalar, which revealed a jump from relevant to irrelevant at the kink location, Fig.~\ref{fig:second-op-bound}. Based on this I argued that the kink should survive in 3D, and even turn into a sharper feature when imposing irrelevance of the second scalar \cite{Scalars2011talk,Rychkov:2011et}.

\begin{figure}[h]
	\centering
	\includegraphics[width=0.45\textwidth]{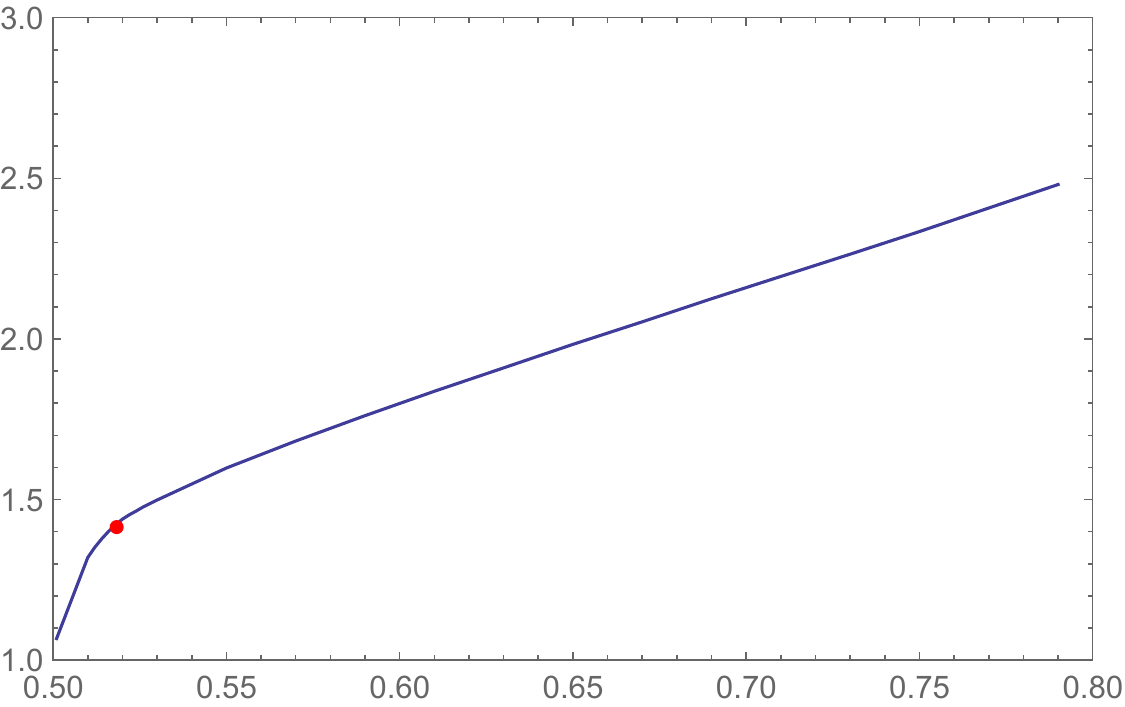}
	\caption{The very first 3D bound, with the red dot at the 3D Ising CFT location expected from the $\eps$-expansion. Email by Alessandro Vichi, November 11, 2011. 
	}
	\label{fig:3d-bound-Vichi}
\end{figure}

\begin{figure}[h]
	\centering
	\includegraphics[width=0.45\textwidth]{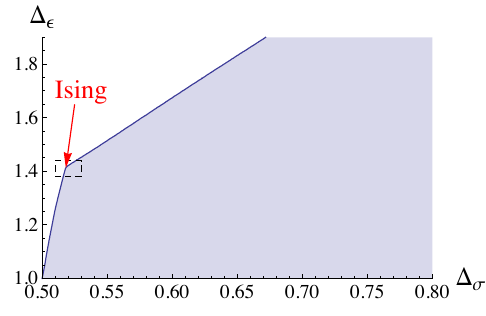}
	\caption{The 3D bound from the paper \cite{El-Showk:2012cjh}.}
	\label{fig:3d-bound-paper}
\end{figure}

That the 3D Ising CFT could thus be located was a tantalizing possibility.
But to check this, one had to find efficient ways to compute 3D conformal blocks and their derivatives. I gave a second talk about this idea in Paris in September 2011, and started collaborating with Miguel Paulos, then a postdoc in Paris. After a few weeks I got in touch with David Poland, as we earlier agreed to inform each other if the 3D ball got rolling. David then suggested to also invite David Simmons-Duffin and Alessandro Vichi who he heard were also getting interested in 3D. On October 13 we had the first group Skype call. Sheer El-Showk, another Parisian postdoc, joined as well, and the 3D Ising collaboration was complete. Most of us knew each other from prior joint works \cite{El-Showk:2011xbs,Costa:2011mg,Costa:2011dw,Poland:2011ey}, and with so much brainpower things started moving rapidly.  We agreed that we would use recursions from Dolan and Osborn \cite{Dolan:2011dv} to reduce the computation to blocks of spins $\ell=0,1$. 
At $z=\bar z$, David Poland and Miguel Paulos found expressions for the $\ell=0,1$ blocks in terms of ${}_3 F_2$'s. For derivatives orthogonal to the $z=\bar z$ line we'd use recursions from the Casimir equation. By the end of October, all key ideas for the 3D blocks were in place, and we could move to a computer implementation. On November 11, Alessandro sent us a Mathematica notebook with the first 3D bound plot, which did show something like a kink at the 3D Ising position (Fig.~\ref{fig:3d-bound-Vichi}). In the paper \cite{El-Showk:2012cjh}, which took a few more month, the kink was sharpened significantly by pushing the derivative expansion order, Fig.~\ref{fig:3d-bound-paper}.

Given the iconic status of the 3D Ising model, our work got noticed. We were encouraged, but there was also the burden of responsibility. Can our results be systematically improved? In September 2012, once our paper was accepted by Phys.Rev.D, we were at the crossroads. 

The most natural directions would be to add more four-point functions. All results in \cite{El-Showk:2012cjh} were based on crossing for $\langle\sigma\sigma\sigma\sigma\rangle$, but in the conclusions of \cite{El-Showk:2012cjh}  we speculated that adding  $\langle\sigma\epsilon\sigma\epsilon\rangle$ and $\langle\epsilon\epsilon\epsilon\epsilon\rangle$ should help constraining the CFT. We could not be sure however. In fact we did try to include $\langle\epsilon\epsilon\epsilon\epsilon\rangle$, but we did not see any improvement. In August 2012, I also tested in 2D a subset of channels of $\langle\sigma\epsilon\sigma\epsilon\rangle$ with positive expansion coefficients $\lambda_{\sigma\epsilon\mathcal{O}}^2$. Again, no improvement. There was still a chance for an improvement including the nonpositive channels of $\langle\sigma\epsilon\sigma\epsilon\rangle$ with coefficients $\lambda_{\sigma\sigma\mathcal{O}}\lambda_{\epsilon\epsilon\mathcal{O}}$. This, however, required switching from linear to semidefinite programming. A semidefinite programming solver was already used for bootstrap studies in $d=4$ in \cite{Poland:2011ey}, 
but interfacing it with our methods for computing 3D blocks seemed nontrivial. This looked like a tough project, with uncertain chances of success, and it was not immediately pursued (but see below).

In the meantime, Miguel and Sheer found that our bounds from \cite{El-Showk:2012cjh} contained more information than visible to the naked eye. By going very closely to the bound they could extract the extremal solution, in which many low-lying exchanged CFT operators stabilized to reasonable accuracy, with their OPE coefficients. While expressed before (\cite{Poland:2010wg}), some of us had doubts how practical this idea could be. Now we were convinced. Miguel and Sheer showcased this "Extremal Functional Method" by reproducing the 2D Ising CFT spectrum \cite{El-Showk:2012vjm}. We could also apply it in 3D if a criterion to fix $\Delta_\sigma$ were found, and rapid changes in the bounds on $\epsilon'$ and $T'$ that we saw in \cite{El-Showk:2012cjh} could help in this task. At the end of September, this was chosen as the primary direction for the collaboration to pursue. 

The hope was that it would take us a few months, but the project took two years of hard work. We ran into limitations of the \emph{machine precision} floating point arithmetic used by the available linear programming solvers. Once agreed that there was no simple way around this, we set out developing our own \emph{arbitrary precision} solvers. 
Eventually we had two of them - a C++ one by David Simmons-Duffin, and a Python one by Sheer El-Showk and myself, both using a primal version of the simplex algorithm with a continuous representation of conformal block derivatives as a function of $\Delta$. In addition, Sheer proposed to replace the $\Delta_\epsilon$-maximization by the $c$-minimization. The latter required a single run of the algorithm, while the former in those days required bisection, hence many runs at every $\Delta_\sigma$.\footnote{Nowadays with the navigator function \cite{Reehorst:2021ykw} $\Delta_\epsilon$-maximization can be also solved in a single run.} With these tools and ideas, our 2014 paper \cite{El-Showk:2014dwa} pushed the conformal bootstrap determination of the main 3D Ising critical exponents $\eta,\nu,\omega$, related to the dimensions of $\sigma,\epsilon,\epsilon'$, factor 3 better than the best Monte Carlo then available. But, as Leo Kadanoff noticed in his appreciative comment \cite{Kadanoff2014}:

 \begin{mybox}
Numerical accuracy is not the main virtue of this paper. This accuracy is
the proof of the pudding. The nourishment, however, is that the numerical
results are obtained from deep understanding of the structure of the Ising
problem. 
 \end{mybox}
 
 \begin{figure}[h]
 	\centering
 	\includegraphics[width=0.45\textwidth]{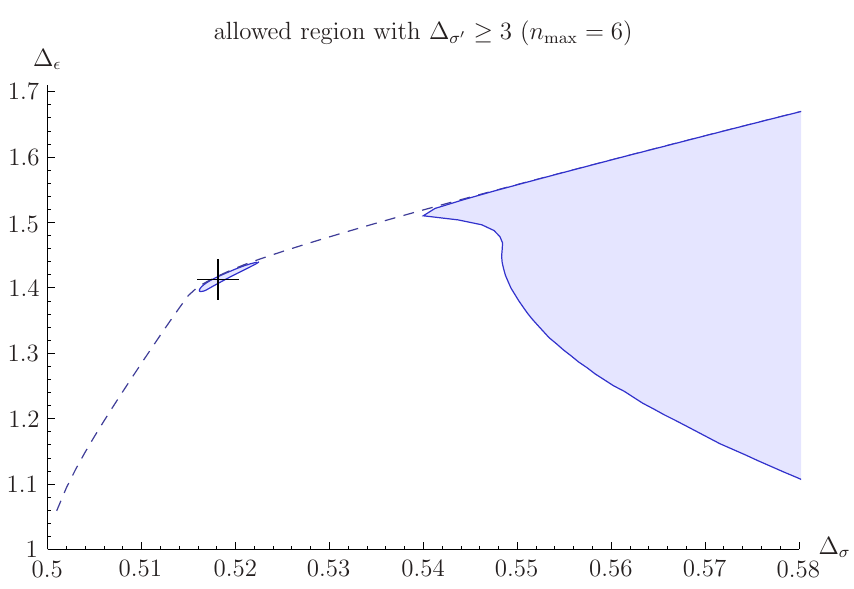}
 	\caption{The first 3D Ising island, from \cite{Kos:2014bka}, surrounded by an excluded region. 
 	}
 	\label{fig:SDP-bound}
 \end{figure}
 
The $\Delta_\epsilon$-maximization \cite{El-Showk:2012cjh} and $c$-minimization \cite{El-Showk:2014dwa} were historically important as the first successful bootstrap attacks on 3D Ising CFT. But ironically it is the above-mentioned idea of adding more four-point correlation functions to the mix that ended up the most powerful. When this was finally implemented, semidefinite programming and all, by Filip Kos, David Poland and David Simmons-Duffin \cite{Kos:2014bka}, they found a spectacular result - a closed allowed region (an "island") - under the assumption of $\sigma$ and $\epsilon$ being the only relevant $\mathbb{Z}_2$-odd and $\mathbb{Z}_2$-even scalars of the 3D Ising CFT (Fig.~\ref{fig:SDP-bound}). This is a more natural assumption than that of the 3D Ising CFT living at the kink used in \cite{El-Showk:2012cjh,El-Showk:2014dwa}. Indeed, it is not an assumption at all but may be taken as a definition of the theory. With subsequent developments, notably introduction of SDPB \cite{Simmons-Duffin:2015qma}, the mixed correlator approach beat all other method, achieving $10^{-5}$ accuracy in the main critical exponents \cite{Kos:2016ysd} and leading to the determination of over a hundred of exchanged operators of the 3D Ising CFT \cite{Simmons-Duffin:2016wlq}. When the stress tensor correlators were recently added to the mix, the accuracy was further improved to $10^{-7}$-$10^{-8}$ \cite{Chang:2024whx}.

\section{2025 - Selected challenges for the conformal bootstrap}\label{sec:2025---selected-challenges-for-the-conformal-bootstrap}

In the decade after 2014, there were many important developments in the conformal bootstrap, as reviewed in \cite{Poland:2016chs,Poland:2018epd,Bissi:2022mrs,Poland:2022qrs,Hartman:2022zik,Rychkov:2023wsd}. 
Here I would like to discuss instead several open problems.

\subsection{Uniqueness problems}

Experiments suggest the critical 3D Ising CFT should be unique—different uniaxial magnets exhibit the same critical exponents, as do different liquid-vapor critical points. Can we prove this uniqueness using bootstrap methods? Fig.~\ref{fig:SDP-bound} shows, in addition to the 3D Ising island, which has since been significantly reduced \cite{Simmons-Duffin:2016wlq,Chang:2024whx}, a "continent" of allowed points to the right which has not experienced a similar reduction (see e.g.~\cite{Atanasov:2022bpi}). In fact the continent cannot go away because it contains other theories which masquerade like 3D Ising CFT while in fact have a different symmetry. One example are the Gross-Neveu-Yukawa models \cite{Erramilli:2022kgp} whose spatial parity $P$ behaves similarly to the global $\mathbb{Z}_2$ as far as correlators of scalars are concerned. These models are excluded when the stress tensor is included in the mix, because $T\times T\not\ni\sigma$ in Ising while $T\times T\ni\sigma$ in GNY (with a parity-odd tensor structure). The continent then shrinks significantly.\footnote{We thank Rajeev Erramilli for communicating to us this unpublished result.} It would be interesting to find further ways to eliminate masquerading theories from the range $\Delta_\sigma,\Delta_\eps<3$, and thus prove the uniqueness of the 3D Ising CFT, defined as the unitary $\mathbb{Z}_2$-invariant theory with a single relevant $\mathbb{Z}_2$-odd and a single relevant $\mathbb{Z}_2$-even scalars.

Alternatively, the continent may not completely disappear but fall apart into several islands, each containing a new CFT satisfying the stated requirement. This will mean that the 3D Ising CFT as defined above is not unique, and further parameters are needed to distinguish such theories. One parameter is the central charge $c$, i.e.~the stress tensor two-point function coefficient, which in 3D does not correspond to any extension of the conformal algebra, but may still be used as a rough count of "degrees of freedom". If the additional theories have a much larger $c$ than the "usual" 3D Ising CFT, they will be harder to realize experimentally, explaining why they have not (yet?) been seen.\footnote{We thank Jo\~ao Penedones for emphasizing this possibility.}

Analogous uniqueness problem may be formulated and studied for other universality classes. 

\subsection{Nonexistence problems}

For some systems, experiments and Monte Carlo simulations indicate first-order transitions. Are we sure that the transition cannot turn second-order by changing the microscopic model a bit. A proof can be obtained by showing that there is no CFT with requisite properties. This is a bootstrap problem.

One example is the 3-state Potts model in 3D. For the nearest-neighbor 3-state Potts model on the 3D cubic lattice, Monte Carlo simulations indicate a first-order transition with a correlation length $\xi\sim 10$ \cite{Janke:1996qb}. The nonexistence bootstrap problem would be then to show that there exists no unitary 3D CFT with $S_3$ global symmetry, one relevant singlet scalar $\eps$ and two relevant scalars $\sigma,\sigma'$ in the fundamental representation.\footnote{The latter counting is by analogy with the 2D 3-state Potts CFT.} This problem is still open although there was interesting recent work \cite{Chester:2022hzt}. For a more detailed discussion, and other examples of nonexistence problems, see \cite{OpenProblems}.

\subsection{Bootstrapping 3D conformal gauge theories}

By the bosonic or ferminic QED${}_3$ CFT we mean the IR fixed point of the 3D $U(1)$ Maxwell theory coupled to $N_f$ bosons or fermions in the UV. The global symmetry of this theory is $SU(N_f)\times U(1)_{\rm top}$, where $U(1)_{\rm top}$ is an emergent "topological" global symmetry whose charge is the magnetic charge of the local monopole operators. These theories are important for contemporary condensed matter physics. For example, bosonic QED${}_3$ with $N_f=2$ is related to the physics of Deconfined Quantum Critical Points, while fermionic QED${}_3$ with $N_f=4$ is related to Dirac Spin Liquids, with a chance to describe a conformal phase of matter in real materials such as the herbertsmithite. This is reviewed e.g.~in \cite[Sec.~V.E]{Poland:2018epd}. It is believed that these theories are conformal for $N_f\ge N_f^*$. An outstanding problem is to determine the critical value $N_f^*$ for the bosonic and fermionic QED${}_3$, as well as the conformal data as a function of $N$.

One class of physical operators of QED${}_3$ are gauge invariant combinations of the elementary fields appearing in the Lagrangian, such as $\bar \psi_i \psi_j$ or $\phi_i^* \phi_j$, as opposed to $\psi_i$, $\phi_i$ themselves which are not gauge invariant. Therefore, these operators are "heavier" (i.e.~have a higher scaling dimension) then the lightest scalar/fermion operators in CFTs without gauge interactions. The monopole operators charged under $U(1)_{\rm top}$ form another class of physical QED${}_3$ operators, which are also heavy, their scaling dimension being $\sim N_f$ in the large $N_f$ limit. 

One difficulty in bootstrapping QED${}_3$ is that, because of larger scaling dimensions $\Delta$, we expect slower convergence as the derivative expansion order $\Lambda$ is increased (see Section \ref{sec:largeDelta} below). We may also expect another difficulty: how shall we distinguish $\text{QED}_3$ from related $\text{QCD}_3$ theories where the gauge group is nonabelian?

While there was a lot of work trying to bootstrap $\text{QED}_3$, and many bounds were derived, because of the above difficulties these theories have not yet been isolated into small closed regions. See \cite{Rychkov:2023wsd} for a review. This remains an interesting open problem. In this context, we would like to mention the recent works \cite{Reehorst:2020phk,He:2021xvg} which identified gaps in the operator spectrum ("decoupling operators") 
in $\text{QED}_3$ which could help distinguish it from $\text{QCD}_3$, where color indices allow for more antisymmetrization, hence more nonvanishing operators to be constructed from elementary fields.

\subsection{Large $\Delta$ problem}
 \label{sec:largeDelta}

\begin{figure}[h]
	\centering
	\includegraphics[width=0.45\textwidth]{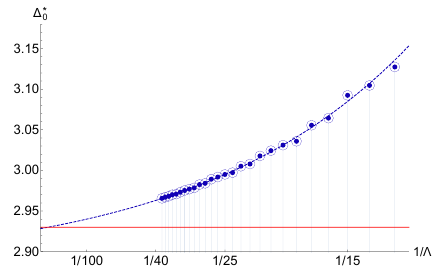}
	\caption{Upper bound on the first unprotected scalar for $\mathcal{N}=4$ SCFT with $c=3/4$ (the would-be Konishi operator in the $\mathcal{N}=4$  SYM with the $SU(2)$ gauge group). This bound was derived from crossing for the four-point function of the protected scalar in $\textbf{20}'$ irrep of the R-symmetry group, of dimension $\Delta=2$. The bound converges rather slowly with $\Lambda$, so from this perspective already $\Delta=2$ is ``large'' \cite{Beem:2016wfs}.\label{fig:Lambda-conv}} 
\end{figure}

This concerns the rate of convergence of bootstrap bounds with increasing order $\Lambda$ of derivative expansion (defined in Eq.~\eqref{eq:derfunc} below). Let $\Delta$ be the scaling dimensions of operators in the four-point functions whose crossing constraints one studies. For small $\Delta$, like $\Delta\approx 0.518$ for the 3D Ising, convergence is rapid, but for larger $\Delta$ it becomes slower. One example is in Fig.~\ref{fig:Lambda-conv}.

\begin{figure}[h]
	\centering
	\includegraphics[width=0.65\textwidth]{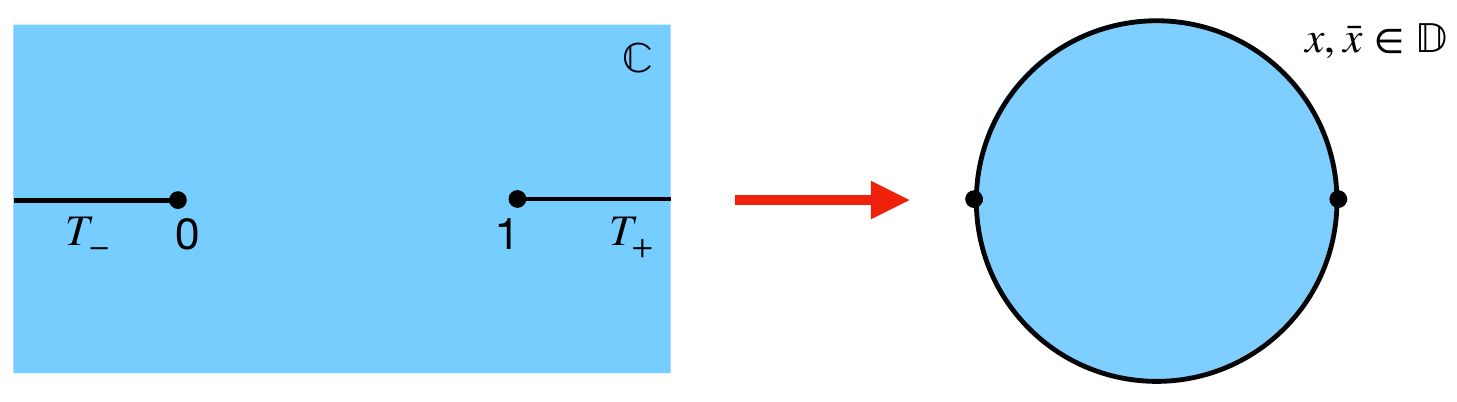}
	\caption{Analytic structure of the four-point function in the $z,\bar z$ variables (left) and after transforming to the $x,\bar x$ variables (right). \label{fig:cuts}} 
\end{figure}

How can we rationalize this problem? Here's some relevant preliminary information. The CFT four-point function $\langle \mathcal{O}_\Delta(0)
\mathcal{O}_\Delta(z,\bar z)\mathcal{O}_\Delta(1) \mathcal{O}_\Delta(\infty)\rangle$ is analytic in $z,\bar z\in \mathbb{C}\backslash (T_+\cup T_-)$ where $T_\pm$ are the two cuts shown in Fig.~\ref{fig:cuts} (left). It has a power singularity $1/(z\bar z)^\Delta$ as $z,\bar z\to 0$, and a symmetric one as $z,\bar z\to 1$. Following \cite{Mazac:2016qev}, let us map the complex plane with two cuts to the unit disk variables $x,\bar x\in \mathbb{D}$,
\beq
z=\frac{(1+x)^2}{2(1+x^2)},\quad \bar z=\frac{(1+\bar x)^2}{2(1+\bar x^2)}\,.
\eeq
The singularities are mapped to $x,\bar x = \pm1$.

The standard way of analyzing crossing is to Taylor-expand it around the point $x=\bar x=0$ to a finite order. This is equivalently formalized as acting on the crossing equation with the "derivative functionals" 
\beq
\partial_x^n \partial_{\bar x}^m|_{x=\bar x=0},\qquad n+m\leq \Lambda\,.
\label{eq:derfunc}
\eeq
The original formulation \cite{Rattazzi:2008pe} used $z,\bar z$ coordinates around the point $z=\bar z=1/2$ which maps to $x, \bar x=0$. When working numerically at finite $\Lambda$, it does not matter if one works in $z,\bar{z}$ or $x,\bar x$, as the two bases are related by a linear transformation. However, if one wants to understand what happens the $\Lambda\to\infty$ limit as we are trying now, then it's best to work in $x,\bar x$. This was first observed by Maz\'a\v{c} in \cite{Mazac:2016qev}, in the 1D case when there is just one cross ratio. 
\begin{figure}[h]
	\centering
	\includegraphics[width=0.25\textwidth]{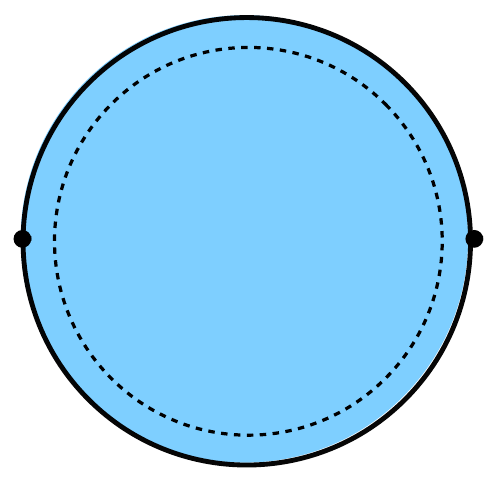}
	\caption{Integrating against a weight along a contour (dashed circle) one gets a linear functional on the space of analytic functions $f(x)$, $x\in \mathbb{D}$. The most general class of functionals is obtained by pushing the contour towards the boundary of analyticity. \label{fig:contour}} 
\end{figure}

Ref.~\cite{Mazac:2016qev} proposed to think of functionals in terms of contour integrals against a weight, pushed to the boundary of the analyticity domain, Fig.~\ref{fig:contour}. The weight provides a more fundamental characterization of the functional than the expansion coefficients in the derivative basis. Some weights correspond to the derivative functionals, others give more general functionals. The values of the function near the boundary of the analyticity domain, on which the functional thus operates, can in principle be reconstructed from the $x,\bar x$ derivatives at the origin, since the power series converges at least in the interior of the domain. 
However the convergence of this power series gets slower with the increase of $\Delta$, which controls the singularity on the boundary. This suggests that:
\begin{itemize}
	\item The extremal functional can be expressed as a series in the $x,\bar x$ derivative functionals. 
	\item The rate of convergence of this expansion may slower as $\Delta$ is increased.
	\end{itemize}
Ref.~\cite{Mazac:2016qev} provided support for these observations, in the 1D case but it's natural to expect that these lessons should retain their usefulness for higher $d$. Furthermore, pursuing this logic, Maz\'a\v{c} \cite{Mazac:2016qev} and Maz\'a\v{c}-Paulos \cite{Mazac:2018mdx,Mazac:2018ycv}, obtained bases of 1D functionals of this "integration over contour" type, having remarkable positivity properties when acting on 1D conformal blocks, dubbed there "analytic functionals". These analytic functionals are nontrivial, slowly convergent linear combinations of derivative functionals.

This not only sheds light on the large $\Delta$ problem, but also suggests a path to its resolution: replace the derivative functionals by the analytic functionals as a basis in the numerical bootstrap.
\begin{figure}[h]
	\centering
	\includegraphics[width=0.6\textwidth]{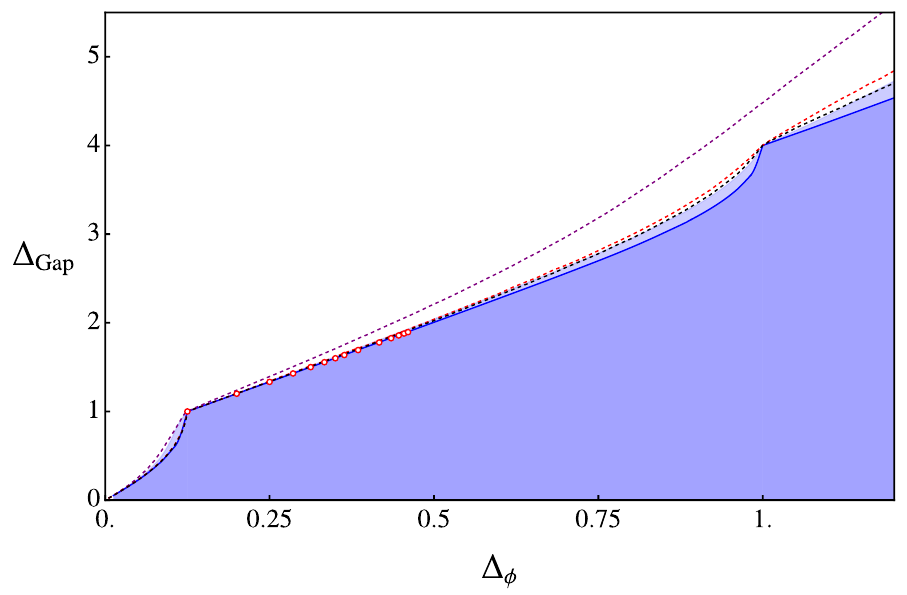}
	\caption{The 2D gap maximization bound analogous to Fig.~\ref{fig:2d-bound}, but obtained in \cite{Ghosh:2023onl} with the higher-$d$ analytic functionals,. It shows very fast convergence compared to the derivative functional basis, especially at higher $\Delta_\phi$. See \cite[Fig.~7]{Ghosh:2023onl} for an explanation.}
	\label{fig:GhoshZheng}
\end{figure}

In the 1D case, this was first done in Ref.~\cite{Paulos:2019fkw}. To extend to the higher-$d$ case, one needs a higher-$d$ basis of analytic functionals with well-defined, computable action on higher-$d$ conformal blocks. Ghosh and Zheng \cite{Ghosh:2023onl} made serious progress toward implementing this program.\footnote{See \cite{Paulos:2019gtx,Mazac:2019shk,Caron-Huot:2020adz} for related prior work.} They use tensor products of 1D analytic functionals in $z$ and $\bar z$ variables. To act on the higher-$d$ blocks, they expand those as a series of 1D blocks of $z$ times 1D blocks of $\bar z$ (the dimensional reduction approach of \cite{Hogervorst:2016hal}). For $d=2$ this series collapses to a finite sum and they get most impressive results, with very fast convergence for high $\Delta$, Fig.~\ref{fig:GhoshZheng}. In $d\ge 3$, to get the action of functionals on conformal blocks, one needs to resum a somewhat slowly convergent series, but they do already have promising preliminary results in $d=3$. Once this is further improved \cite{GhoshZheng}, this method may well become the future of the numerical conformal bootstrap.

\section{Conclusions}

One goal of historical papers is to inspire young researchers. What lessons can they learn from this story? Two times in its history, conformal field theory benefited from the flow of ideas between particle theory and statistical physics. First at its birth, when Polyakov and others were motivated by both hadronic physics and by the critical phenomena. And then forty years later, when our generation started with a sharply defined question from Beyond the Standard Model phenomenology, and finished with the 3D Ising model critical exponents. This is an incredible example of continuing interconnectedness of theoretical physics.

Should we regard conformal symmetry as experimentally confirmed? We have seen that conformal field theory predicts scaling dimensions of operators, and critical exponents are simple functions of those. For example, the 3D Ising model critical exponents $\eta$ and $\nu$ are related to $\Delta_\sigma$ and $\Delta_\epsilon$ by
\beq
\eta = 2\Delta_\sigma-1,\quad \nu =\frac{1}{3-\Delta_\epsilon}\,.
\eeq
This gives results in agreement with measurements in the lab and in Monte Carlo simulations. See Table~\ref{tab:3DIsing}, where we also include the results from the Renormalization Group. Experimental results in the table are a rough summary of the 20-year-old measurements in the liquid-vapor transitions, binary fluids, and uniaxial magnets from Table 7 of \cite{Pelissetto:2000ek}; the accuracy is not outstanding, unfortunately. The RG results are from the 6-loop resummed $\epsilon$-expansion by Kompaniets and Panzer \cite{Kompaniets:2017yct}. Hasenbusch's  improved action Monte Carlo simulations \cite{Hasenbusch:2011yya} are a bit more accurate than RG. Finally, the CFT result in Table~\ref{tab:3DIsing} is from conformal bootstrap analysis using the $\epsilon,\sigma,T$ mix \cite{Chang:2024whx}. It has by far the highest accuracy than any other method, and is consistent with all of them. 
\begin{table}[h]
	\centering
	\begin{tabular}{lll}
		\toprule
		&$\eta$ & $\nu$ \\
		\midrule
		Experiments \cite[Table 7]{Pelissetto:2000ek} &0.04(1) & 0.63(1)\\
		Renormalization Group \cite{Kompaniets:2017yct} & 0.0362(6) &  0.6292(5)\\
		Monte Carlo \cite{Hasenbusch:2011yya}  & 0.03627(10) & 0.63002(10)\\
		CFT \cite{Chang:2024whx} & 0.036297612(48) & 0.62997097(12)\\
		\bottomrule
	\end{tabular}
	\caption{Results for 3D Ising critical exponents from various approaches, including CFT.}\label{tab:3DIsing}
\end{table}

The same argument between CFT, Monte Carlo, RG and experiments holds for many other universality classes, in 2D and in 3D. For example:
\begin{itemize}
	\item
	3D XY class, see \cite{Kos:2016ysd,Chester:2019ifh} for CFT results, and references therein for other methods. In this case the CFT agrees with the RG and the Monte Carlo, but there is a 4th digit disagreement with a very precise liquid Helium experiment, necessitating further experimental study.
	\item
	3D Heisenberg class, see \cite{Kos:2016ysd,Chester:2020iyt} for the CFT results, and references therein for other methods.
	\item
	In 2D, we mention the Ising class and the 3-state Potts class which are both described by exactly solvable minimal model 2D CFTs. For many experimental verifications of the critical exponents, see the references in \cite[pp. 22-27]{Henkel2017_PhaseTransitions2D} as well as \cite{Back1995,WurschPescia1998}.
	\end{itemize}
	
The agreement of CFT with other techniques shows that all these critical points are conformally invariant.\footnote{Not all universality classes are conformally invariant, see e.g.~\cite{Riva:2005gd,Mauri:2021ili,Gimenez-Grau:2023lpz}, but a good fraction of them are. It is well understood which ones are and which ones are not \cite{Polchinski:1987dy,Nakayama:2013is}.}\ \footnote{In 2D, there are also mathematical proofs of conformal invariance of critical points of percolation and of the nearest-neighbor Ising model, using discrete complex analysis (Stanislav Smirnov, Fields Medal, 2010).} Therefore, conformal invariance is a true emergent symmetry of nature. 
Of course, further experimental tests are welcome, including observables sensitive to conformal kinematics such as three- and higher-point correlators or two-point functions in presence of defects and boundaries (see \cite{Podo:2026hfh} for a proposal). Unfortunately, I am not aware of any such more direct test in the lab, apart from the measurements of critical exponents. For some direct tests of conformal invariance in numerical experiments for the 3D Ising model, see \cite{Cosme:2015cxa,Zhu:2022gjc}.

\begin{acknowledgments}
	
	This article is partly based on a Colloquium at the Enrico Fermi Institute, University of Chicago (October 30, 2023), and on a talk at the Wolfgang Pauli Center Theoretical Physics Symposium, DESY Hamburg (May 15, 2025).
	I thank Dam Thanh Son and Volker Schomerus for the kind invitations. I am indebted to my coauthors of \cite{Rattazzi:2008pe,El-Showk:2012cjh} who read a draft of this article and contributed their memories and copies of lost emails. S.R. was supported in part by the Simons Foundation grant 733758 (Simons Bootstrap Collaboration).
	
	\end{acknowledgments}


\providecommand{\href}[2]{#2}\begingroup\raggedright\endgroup

\end{document}